\documentclass[sigconf]{acmart}
\usepackage{booktabs} % For formal tables
\usepackage{xspace}
\usepackage{paralist}
\usepackage{ifthen}
\usepackage[normalem]{ulem} % for \sout
\usepackage{xcolor}

\newboolean{showedits}
\setboolean{showedits}{true} % toggle to show or hide edits
\ifthenelse{\boolean{showedits}}
{
	\newcommand{\del}[1]{\textcolor{red}{\sout{#1}}} % please delete
}{
	\newcommand{\del}[1]{} % please delete
	
}
\newboolean{showcomments}
\setboolean{showcomments}{true}
\newcommand{\id}[1]{$-$Id: scg-llncs.tex 30911 2010-02-05 10:21:47Z oscar $-$}

\ifthenelse{\boolean{showcomments}}
{\newcommand{\nbc}[3]{
 {\colorbox{#3}{\bfseries\sffamily\scriptsize\textcolor{white}{#1}}}
 {\textcolor{#3}{\sf\small$\blacktriangleright$\textit{#2}$\blacktriangleleft$}}}
 }
{\newcommand{\nbc}[3]{}
 \renewcommand{\del}[1]{} % please delete
 }

\newcommand{\ie}{\emph{i.e.},\xspace}
\newcommand{\eg}{\emph{e.g.},\xspace}
\newcommand{\etal}{\emph{et al.}\xspace}

\setcopyright{rightsretained}
\copyrightyear{2019} 
\acmYear{2019} 
\setcopyright{acmlicensed}
\acmConference[ICPE '19 Companion]{Tenth ACM/SPEC International Conference on Performance Engineering Companion}{April 7--11, 2019}{Mumbai, India}
\acmBooktitle{Tenth ACM/SPEC International Conference on Performance Engineering Companion (ICPE '19 Companion), April 7--11, 2019, Mumbai, India}
\acmPrice{15.00}
\acmDOI{10.1145/3302541.3313104}
\acmISBN{978-1-4503-6286-3/19/04}

\begin{document}
\title{PerfVis: Pervasive Visualization in Immersive Augmented Reality for Performance Awareness}

\author{Leonel Merino}
\affiliation{%
  \institution{VISUS, University of Stuttgart}
}
%\email{leonel.merino@visus.uni-stuttgart.de}

\author{Mario Hess}
\affiliation{%
  \institution{SCG, University of Bern}
}
%\email{hessm1993@gmail.com}

\author{Alexandre Bergel}
\affiliation{%
  \institution{ISCLab, DCC, University of Chile}
}
%\email{abergel@dcc.uchile.cl}

\author{Oscar Nierstrasz}
\affiliation{%
  \institution{SCG, University of Bern}
}
%\email{oscar@inf.unibe.ch}

\author{Daniel Weiskopf}
\affiliation{%
  \institution{VISUS, University of Stuttgart}
}
%\email{weiskopf@visus.uni-stuttgart.de}

% The default list of authors is too long for headers.
\renewcommand{\shortauthors}{L. Merino et al.}

\begin{abstract}
Developers are usually unaware of the impact of code changes to the performance of software systems. Although developers can analyze the performance of a system by executing, for instance, a performance test to compare the performance of two consecutive versions of the system, changing from a programming task to a testing task would disrupt the development flow. In this paper, we propose the use of a city visualization that dynamically provides developers with a pervasive view of the continuous performance of a system. We use an immersive augmented reality device (Microsoft HoloLens) to display our visualization and extend the integrated development environment on a computer screen to use the physical space. We report on technical details of the design and implementation of our visualization tool, and discuss early feedback that we collected of its usability. Our investigation explores a new visual metaphor to support the exploration and analysis of possibly very large and multidimensional performance data. Our initial result indicates that the city metaphor can be adequate to analyze dynamic performance data on a large and non-trivial software system.
\end{abstract}

%
% The code below should be generated by the tool at
% http://dl.acm.org/ccs.cfm
% Please copy and paste the code instead of the example below.
%


%\keywords{ACM proceedings, \LaTeX, text tagging}

\maketitle
%****************************%
\section{Introduction}
%****************************%
Among the many questions that arise during software development, programmers often formulate questions about the performance of software systems~\cite{Lato10b,Sand13a}. They ask, for instance, ``what is the software doing when performance issues arise?''~\cite{Reis05a}, and ``where is most of the time being spent?''~\cite{Shah16a}. Since developers introduce multiple changes to the source code during the implementation of a software system, being aware of the impact of a code change task to the performance of a software system is an important concern for developers~\cite{Lato10b}. 
Tools such as profilers and performance tests allow developers to analyze the performance of only a single version of a system at a time, and force developers to change to a different task (interrupting their flow). Our intuition is that developers would benefit from analyzing the live performance of an evolving system as its source code changes. We expect that such an analysis would help them identify changes that severely decrease the performance of a system. 
\begin{figure}[t!] 
	\centering
	\includegraphics[width=\linewidth]{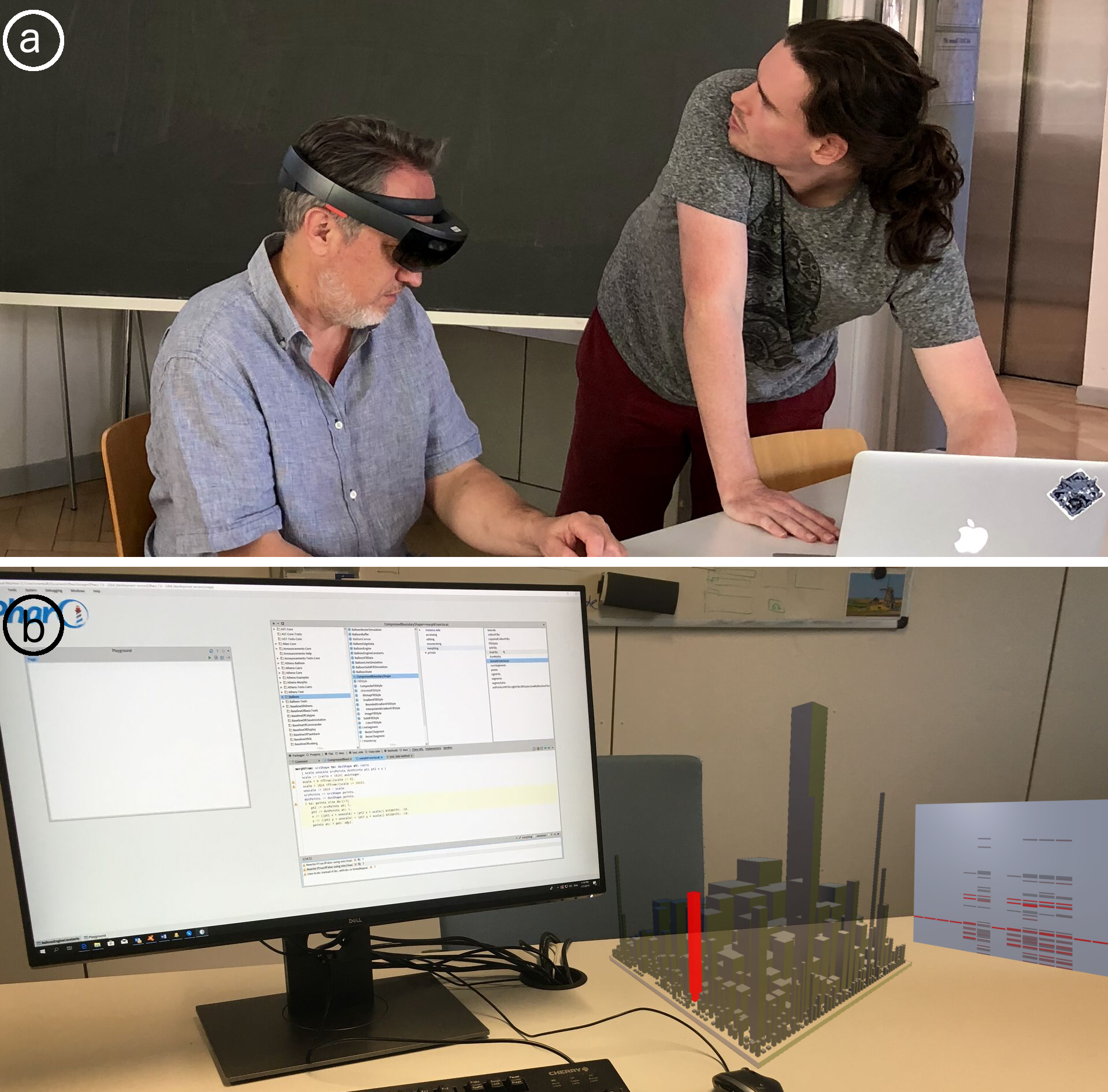}
	\vspace{-1.5em}
	\caption{Pervasive visualization displayed in immersive augmented reality: \textcircled{a} a developer wears a Microsoft HoloLens device \textcircled{b} for performance visualizations using the city metaphor and a scatter plot.}
	\vspace{-1.5em}
	\label{fig:immersive}
\end{figure} 
We conjecture that a pervasive tool, (\ie a tool that is omnipresent at all times during the implementation of a software system) can make developers aware of important changes in the performance of a system without disrupting the implementation flow.
Using a visualization approach, developers could obtain such a pervasive view of the live performance of a system, however, we observe that developers can be reluctant to sacrifice space in their integrated development environment (IDE). Consequently, we formulate the following research question: 
\framebox{
	\parbox[t][0.6cm]{0.95\linewidth}{
	\addvspace{-0.05cm}
        \noindent\emph{RQ: How can visualization support developers in the analysis of the impact of source code changes to the performance of a system?}
	} 

}\\

This paper presents an initial step toward answering our general research question.
To increase developer awareness of the impact of changes to the source code in the performance of software systems we present \emph{PerfVis}, a pervasive visualization tool displaying in immersive augmented reality. PerfVis provides an omnipresent overview of the system that takes advantage of visualization techniques to make developers aware of the impact of changes introduced to the source code in the live performance of a software system. 
%can provide engineers great support to monitor the system’s behavior by enabling them to identify visual patterns that might lead to actual anomalies in the system. However, the main challenge for engineers who are monitoring a running system is to maintain the focus in the tool to not miss data while performing other tasks.

%ProfVis: What parts of the program could be modified to improve performance?~\cite{Lin10a}
%lviz: How the operating system works?~\cite{Wu10a}
%Zinsight: How did we get to these events?~\cite{Paw10a}
%Streamsight: How the system and applications evolve?~\cite{Paw08a}
%MemoView: How does fragmentation depend on time and pool?~\cite{More07a}
%IBM Web Services Navigator tool : How different IT resources interact sequentially with one another?~\cite{Pauw06a}
%DOT: What are optimal parameters to distribute the work on the processors?~\cite{Bloc05a}

The main contribution of the paper is threefold: 
\begin{inparaenum}[(i\upshape)]
    \item a discussion of design choices, 
    \item lessons learned from building the tool, and
    \item insights from observing its use. %~\cite{Meri18a}.
\end{inparaenum}
We also contribute to the reproducibility of our research by making the implementation of PerfVis publicly available\footnote{\url{http://scg.unibe.ch/research/perfvis}}.

%****************************%
\section{Related work}
%****************************%
%We identify three main categories of related work
%\begin{inparaenum}[(\itshape i\upshape)]
%    \item 3D visualizations,
%    \item visualizations displayed in immersive augmented and virtual reality, and
%    \item visualization of software performance.
%\end{inparaenum}
%In the following we elaborate on the main aspects of the related work, and discuss how they differ from our proposed approach.

%\noindent\textbf{[Technique] \emph{3D Visualization}}. 
A number of 3D visualization tools have been proposed to support software development concerns. \emph{SeeIt 3D}~\cite{Shar13a} uses a 3D scatter plot visualization to represent software metrics collected from Java source code to support developers on software comprehension tasks. %A comparative evaluation between the proposed visualization tool with a popular integrated development environment (\ie Eclipse) that included the analysis of data collected from eye tracking showed that participants who used SeeIT 3D achieved a better performance in overview tasks, but took longer time to bug-fix tasks. 
\emph{SynchroVis}~\cite{Wall13a} employs a visualization based on the 3D city metaphor that displays the structure and metrics of software systems, as well as program traces and synchronization aspects (\eg semaphores/wait) to support the analysis of the behavior of concurrent software systems.  
%A 3D visualization tool~\cite{Tymc16a} proposes to identify quality anomalies in dynamically typed software systems by allowing developers to analyze the co-evolution of source code and quality metrics in a software system. The visualization uses the space-time cube technique to represent 
%\begin{inparaenum}[(\itshape i\upshape)]
%    \item software components, 
%    \item software versions, and 
%    \item broken quality rules. 
%\end{inparaenum}
These tools display the 3D visualization on a standard computer screen and focus on performance data collected using static analysis. In contrast, we propose the visualization to be displayed in immersive augmented reality and focus on live performance data collected using dynamic analysis.  

There are some visualization tools that use virtual reality to support software comprehension tasks. \emph{Code Park}~\cite{Khal17a} proposes an immersive visualization of software metrics and source code; an evaluation showed that the visualization displayed in virtual reality excels at usability and significantly helps in code understanding. \emph{ExplorViz}~\cite{Fitt17a} introduces a visualization of software landscapes to support software comprehension tasks. \emph{CityVR}~\cite{Meri17c} gamifies software implementations using a 3D city visualization displayed in virtual reality to boost developer engagement. %The tool visualizes the static structure of software using the city metaphor.
%Early results of developers using the visualization show benefits engagement such as promoting curiosity, immersion, and excitement, and making developers willing to spend considerable time using the visualization.
In contrast, we propose the use of immersive augmented reality to improve the awareness of the impact of changes on the source code to the performance of software systems. Moreover, some of these tools focus on various static software metrics, as opposed to our tool that includes the visualization of dynamic live performance data.

%\noindent\textbf{[Task] \emph{Visualization of Performance}.}
%The visualization of live performance is challenging due the short time span that users have to analyze the data. In consequence, 
Several existing performance visualizations focus on the analysis of static performance data. \emph{JOVE}~\cite{Reis05a} is a visualization tool for monitoring the performance of Java programs. %To obtain an omnipresent view of the performance of the system, the tool that adds little overhead to the programming environment. 
Using the visualization, developers can obtain details on demand to identify routines and threads in which the application spends much time.  
Another tool~\cite{Bloc05a} includes the visualization of the structure of parallel software systems. The visualization uses an execution graph to simplify the analysis of the complex run-time behavior. 
Moreta and Telea~\cite{More07a} present a visualization of the behavior of the memory allocator in C programs to optimize functionality, decrease fragmentation, and improve response time. %The visualization uses a \ugh{Cartesian layout} is used to show how the memory usage changes over time. 
Another tool~\cite{Ogam17} facilitates the visualization of software performance in real-time, using the city metaphor to show the structure and performance of a program. The buildings in the city represent the classes in the system, and the heights of the buildings represent the number of times the methods of a class are called during execution. 
All these tools support the analysis of various aspects of software performance through visualizations displayed on a standard computer screen. In contrast, we rely on immersive augmented reality to display our visualization.

A few software visualizations chose augmented reality as a display medium. One study~\cite{Meri18c} evaluated a city visualization displayed in immersive augmented reality to support software comprehension tasks. Another study~\cite{Souz12a} used the city metaphor in augmented reality to support the analysis of software evolution. In contrast, we propose  a city-based visualization using immersive augmented reality to support software performance tasks, which is an innovative and not-yet explored approach. 

%****************************%
\section{PerfVis Overview}
%****************************%
PerfVis is a tool to visualize software performance through immersive augmented reality. 
%Using PerfVis  developers can be aware of the workload of a software system without requiring their constant attention. 
%To summarize 
We characterize the context of PerfVis using a taxonomy proposed by Maletic \etal~\cite{Male02a} and Merino \etal~\cite{Meri17a}.
\framebox{
    \vspace{1.5em}
	\parbox[t][1.4cm]{0.97\linewidth}{
	\addvspace{-0.05cm}
        \emph{PerfVis} provides visualization in \emph{immersive augmented reality} to support the \emph{programmer} audience in system performance tasks. The visualization shows \emph{static data} of the structure of the system and \emph{dynamic data} of the live performance of the system. 
	} 
}\\
\vspace{-0.5em}	
\begin{figure}[t!] 
	\centering
	\includegraphics[width=0.98\linewidth]{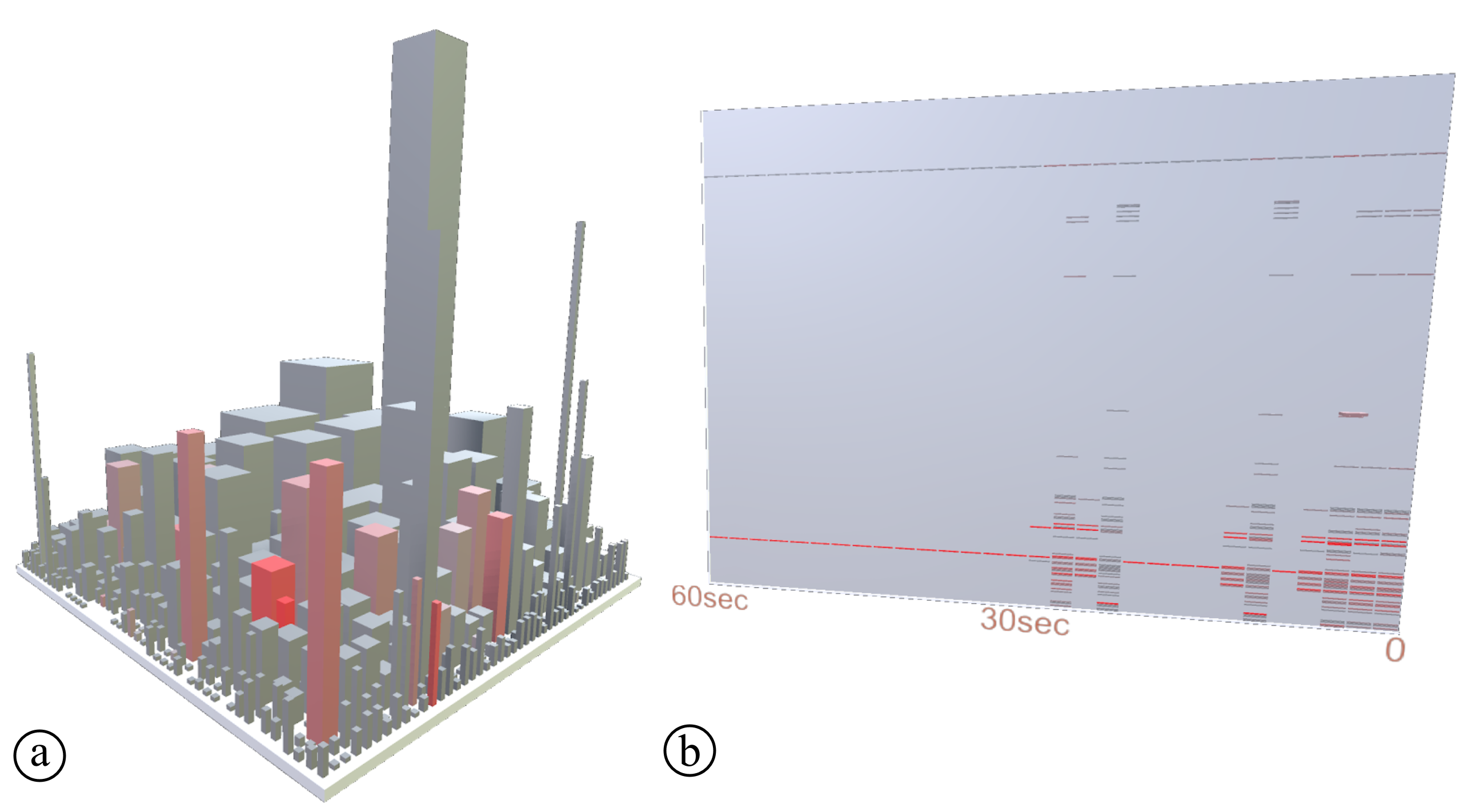}
	\vspace{-1.5em}
	\caption{A detailed view of the \textcircled{a} dynamic city visualization and \textcircled{b} the complementary scatter plot that maintains the history of the overall performance of the system for a configurable interval of time.}
	\vspace{-1.3em}
	\label{fig:vis}
\end{figure} 
    \subsection{Design}
    
        \textbf{Medium: Immersive Augmented Reality.}
            Most visualizations that support performance analysis tasks are displayed on a computer screen. They require developers to either use an application outside their development environment, or use a plug-in in their environment that sacrifices some valuable screen space. Amongst the few visualizations that use a different medium most are displayed in immersive virtual reality~\cite{Meri17b}. These visualizations require that developers completely change their development environment, and isolate themselves wearing a headset. As opposed to these approaches, we argue that a medium that disrupts as little as possible the development flow can significantly increase the usability of visualizations. The medium used to display a software visualization can impact its effectiveness and efficiency~\cite{Meri18b}. Consequently, we chose immersive augmented reality to display PerfVis, because 
            \begin{inparaenum}[(i\upshape)]
            \item 
                developers can use the visualization as a complement to their usual development environment without much disruption (\ie developers are required to wear a headset, but not to modify the rest of their environment), and
            \item visualizations in immersive augmented reality are inherently displayed in 3D, which provides an extra dimension to encode performance data of a software system.   
            \end{inparaenum}

        \noindent\textbf{Apparatus: HoloLens}.
        We choose Microsoft HoloLens to render our visualization in immersive augmented reality. HoloLens is an untethered head-mounted stereoscopic display, featuring two stereoscopic displays with 1268$\times$720 pixel resolution, 60 Hz content refresh rate, and a 30\textdegree~H and 17.5\textdegree~V field of view. Interaction is possible through head tracking, gesture input, and voice support.
        
        \noindent\textbf{Technique: City Visualization and Scatter plot.}
            %Although 3D visualization are widely accepted in scientific visualization \todo{Add citations}, the visualization of abstract data (\ie data that do not have an inherent spatial representation) is controversial. 
            %A number of usability issues of 3D visualizations have been discussed~\cite{Brat14a} (\eg navigation, occlusion, manipulation and text reading). A few studies have addressed the analysis of the reported usability issues. 
            %One study~\cite{Wood05c} investigated the relation between the dimensions of the data and the medium used to display a visualization technique across the various stages of the visualization process. The study confirms that 3D visualizations can be expressive, effective and appropriate, when they provide depth clues (\eg perspective, motion) that ease navigation. 
            In a previous study~\cite{Meri17a}, we observed that most visualizations to support developers in monitoring performance used pixel-based techniques. These visualizations usually attempt to provide an overview of a significant amount of data to promote the detection of visual patterns that may signal anomalies in the performance of software systems. Instead, we adopt lightweight visualization techniques that involve less cognitive effort to make developers aware of the performance of a system. Our visualization includes two views:
            \begin{inparaenum}[(i\upshape)]
                \item a city visualization that provides a pervasive overview of the structure and performance and 
                \item a scatter plot that displays the history of the performance of a software system in the past.
            \end{inparaenum}

            %We chose the city metaphor not only because it has shown to be effective in system comprehension tasks but also because of the familiarity of the city metaphor. %Since the visualization is intended as an omnipresent companion to an integrated development environment, it has to offer useful information, without overwhelming the user with its complexity. 
            %We argue that in order to increase the usability of PerfVis, we limited the complexity of the chosen visualization techniques. 
            
            \emph{City visualization.}
            We opted to represent a metric of the performance of the software system using the color of buildings. The visualization is always on and indicates the system performance collected during a short time frame.
            In consequence, a glimpse at the city can make developers aware of the current overall status of the performance, and a detail of the classes that are involved. %In Figure \ref{fig:vis}, buildings are colored according to the workload of the representing class. 
            Figure~\ref{fig:vis}~{\large \textcircled{\small a}} shows a visualized software city. Each building represents a class in the software system. Buildings are grouped into districts according to their packages. Each building can be configured to represent software metrics using three properties:
            \begin{inparaenum}[(i\upshape)]
                \item height,
                \item footprint (\eg squared base), and
                \item color.
            \end{inparaenum}
            We consequently configured the city in the following manner: each building's 
            \begin{inparaenum}[(i\upshape)]
                \item height encodes the number of methods of the represented class,
                \item footprint (\ie squared base) encodes the number of attributes  of the represented class, and 
                \item color encodes the number of times that methods of the class are called during execution.
            \end{inparaenum}
            
            \emph{Scatter plot.}
            The scatter plot helps developers analyze changes in the dynamic city visualization that might be overlooked. %An overview of the immediate past performance can help alleviate this problem. %But developers might, for other reasons, be interested in the immediate history. For example to compare the performance at different times, search for outliers or inspect the behaviour over time.
            Figure~\ref{fig:vis}~{\large \textcircled{\small b}} shows the scatter plot to the right of the city visualization. The $X$-axis encodes time, and the $Y$-axis encodes the classes in the system. Each mark represents a class that was involved in the execution of the system at a point in time. The color of a mark encodes the number of times methods of a class were called. The more intense the red, the more often a class was called. Each row in the scatter plot represents the evolution of the number of times the methods of a class are called along time. Each column is a snapshot of the performance of the whole system at a point in time. %A horizontal cut through the scatter plot reveals the behaviour of one class over time. A vertical cut, reveals the performance of the entire software at some point in the past.
        
        \noindent\textbf{Interaction: Selection and Navigation}.
            Developers can interact with PerfVis through: %The visualization is scaled to fit the desktop of developers, but not obstruct the existing setup. The goal is to provide an overview from a glance, and allow more information on further investigation. The user can investigate an interesting class by looking at it directly. 
            \begin{inparaenum}[(i\upshape)]
                \item \emph{Selection}. Through head movements, users can point and select elements in the visualization. When the pointer hovers over a building in the city visualization or a mark in the scatter plot, the name of the corresponding class is displayed. %That way, the user can observe static and dynamic metrics visualized by the building. 
                Users select buildings by performing an \emph{airtap} gesture.\footnote{\url{https://docs.microsoft.com/en-us/hololens/}} Once a building is selected, that building and the mark that represent the same class are simultaneously highlighted with a yellow background. Similarly, when selecting a mark in the scatter plot, the building and mark that represent the same class are highlighted. Using this feature, developers relate the information they get from one visualization to the elements in the complementary visualization. %from This way, the selected class is quickly re-identified when the user's focus is interrupted, and the history of the selected building can be inspected in the scatter plot. When studying the scatter plot, gazing at an element will display the name of the corresponding class. To learn which building corresponds to a horizontal line, the user can select a line, which highlights said line and also the corresponding building.
                \item \emph{Navigation}. To gain an overview of the whole system, the user can move around and inspect the city from different angles. This way the user can compare metrics between different buildings. 
            \end{inparaenum}

\subsection{Implementation} The landscape of the city visualization is generated using CodeCity\footnote{\url{http://smalltalkhub.com/\#!/~RichardWettel/CodeCity}} for Moose 5.\footnote{\url{http://www.moosetechnology.org/}}
The color animation of the city and the scatter plot are implemented in Unity\footnote{Version 2017.3.0b8 (64-bit)} using the MixedRealityToolkit-Unity\footnote{\url{https://github.com/Microsoft/MixedRealityToolkit-Unity}} library. The performance data are collected using the Spy2\footnote{\url{https://github.com/ObjectProfile/Spy2}} profiler for Pharo Smalltalk, and streamed to the HoloLens through the network.

\section{Discussion}
%****************************%
We revisit our research question:
\emph{How can visualization support developers in the analysis of the impact of source code changes to the performance of a system?}
We have presented a design and publicly available implementation of a visualization tool that uses immersive augmented reality to support developers in the analysis of software performance. We now discuss lessons learned during the development process, and early feedback on the usability of PerfVis. 

    \subsection{Lessons Learned}
        \noindent\textbf{Scale.} PerfVis aims to give developers an overview of the live performance of a software system with the least disruption possible to the integrated development environment. We tested PerfVis visualizations at various scales. We observed that the visualization had to be scaled to be large enough to give an overview of the complete system, but also small enough to be used as a complement to a computer screen.%seen peripheral vision.
        %With PerVis we tried to give developers a lightweight visualization of performance that does not require constant focus but still provides an overview over the whole system. We learned that the scale  plays a vital role in usability and usefulness of the visualization.  
        
        %\lm{mention that we tested the visualization scaled to various sizes. We opted the final scale because it is big enough to provide an overview of the entire system, but small enough to see it with peripheral vision.}
        
        \noindent\textbf{Text.} %After catching some anomaly in the visualization, the next step in understanding it, is to dive deeper. In our approach, this involves finding out the name of the corresponding class of a building and exploring its history. In a standard computer screen, display and reading 
        %Several properties have to be considered when presenting text in immersive augmented reality, \eg position, rotation, scale, and contrast. 
        We learned that the position, rotation, and scale of text can greatly impact usability when navigation is done by head movements. In an early prototype, we chose to display the names of classes on top of their corresponding building, however, we found that depending on the position of the user and the size of the building, the text on top of a building might be difficult to read, and also that text on top of a building might be occluded by neighboring buildings. We addressed these issues by positioning the text in relation to the position of the camera. Specifically, we placed the text at the top left corner of the user's field of view. Benefits of this approach are that 
        \begin{inparaenum}[(i\upshape)]
            \item text is never occluded, 
            \item text always faces the user, and 
            \item no movement of the head and only little movement of the eyes is required to read the text.
         \end{inparaenum}

        \noindent\textbf{History.} A live visualization has the benefit that it can provide developers online feedback of the performance of the system, however, changes in the visual properties of the visualization can take a very short time. Then the city visualization can be good at making developers aware of an important change in the performance but might not be good for analyzing details of the context of the change since it provides developers too little time to react. To help developers get details on demand we included a companion lightweight visualization (\ie scatter plot). 
        %We decided to add the scatter plot as a holographic computer screen (\ie a rectangular area) to the visualization. 
        We observed some usability issues. For instance, 
        \begin{inparaenum}[(i\upshape)]
            \item marks in the scatter plot can be too small and difficult to select, and
            \item the size of the buffer presented in the scatter plot has to ensure a good performance (too many marks would slow down the visualization tool).
        \end{inparaenum}
        To address the former issue we included a type of selection that involves only head movement (marks are selected when hovering over them for one second). We also addressed the latter issue by making the definition of the size of the buffer configurable. 
%\vspace{-0.3em}      
        %\lm{discuss the implications of this choice, and ways to improve it.}
    \subsection{Early Feedback on Usability}
    %To obtain early feedback that allows us to improve our visualization tool, we invited a few developers to use PerfVis. Additionally, w
    We demonstrated the tool to a group of Computer Science students to discuss our design choices and obtain impressions. Their feedback allowed us to iterate several times on the design of the various components of the visualization tool. Certainly, the positive impressions of developers who used our visualization tool do not represent a solid proof of its benefits, however, they represent a basis to design a suitable evaluation~\cite{Meri18a}.% In the following we discuss some of the impressions that we obtained during this preliminary evaluation.

    \noindent\textbf{Visualization Techniques.} Developers who used PerfVis noticed that the city visualization and the scatter plot complement each other. The former, being a space-filling technique, can be scaled down to a small size, and so provide developers an omnipresent overview of the performance of the whole system; and once the attention of a developer is caught, the latter can be used to provide details on demand of the history of the performance of a system. Some developers who used early prototypes of PerfVis observed that the city visualization was too big, and so only part of it was visible through the narrow field of view of the immersive augmented reality device that we used (HoloLens). We then scaled the visualization to be visible even when the developer is looking at the computer screen.
    
    \noindent\textbf{Interactions.} Interactions such as navigation and selection represent key features that promote usability. In particular, developers who used PerfVis found it difficult to stabilize their head when pointing at small elements in the visualization to select them (through airtap). Some of the developers also said that the highlighting feature of selected elements in the city and scatter plot visualizations did not have a good contrast, and it was hard to see. Several developers suggested including more interactions to the scatter plot visualization. Specifically, they said that features to pause and rewind would help them in the analysis of the performance of the system in a past time.

   % \noindent\textbf{Emotions.} Many emotions arise when developers use PerfVis. Most developers found \emph{easy} to navigate the visualizations. Some developers felt that using the visualization tool resembled playing a computer game. Some other developers felt confident to identify the classes that are used. Several developers said that the city visualization facilitates making estimations on-the-fly. Developers felt many positive feelings, and observed that the immersive augmented reality medium can improve the usability of tools that support tedious and laborious tasks (\eg debugging). A few negative emotions also arose. A few developers felt pain in the nose after wearing for approximately 30 minutes the MS HoloLens headset. They observed that the headset might be too heavy for using it for long periods of time.

%****************************%
\section{Conclusion and Future work}
%****************************%
We introduced \emph{PerfVis}, a visualization tool displayed in immersive augmented reality that supports developers in the analysis of software performance. We elaborated on design choices and discussed implementation details, lessons learned, and early impressions of developers who used prototypes of our visualization tool. In the future, we plan to conduct a thorough evaluation of PerfVis to identify strengths and improvement opportunities.

\section*{Acknowledgments}
Merino and Weiskopf acknowledge funding by the Deutsche For\-schungsgemeinschaft (DFG, German Research Foundation) -- Projektnummer 251654672 -- TRR 161.
Nierstrasz thanks the Swiss National Science Foundation for its financial support of ``Agile Software Analysis'' (project 162352). 
Bergel thanks LAM Research for its financial support.

\bibliographystyle{ACM-Reference-Format}
\bibliography{perfvis}

\end{document}